\documentclass[twocolumn,showpacs]{revtex4}
\usepackage{graphicx}

\begin{document}
\title{First-order Synchronization Transition in Locally Coupled Maps}
\author{ P. K. Mohanty}
\affiliation {Department of Physics of Complex Systems, Weizmann 
Institute of Science, Rehovot, Israel 76100.}
\date{\today}

%%%%%%%%%%%%%%%-------NEW COMMAND-------------------%%%%%%%%%%%%%%%%
\newcommand{\cl}{\centerline}
\newcommand{\be}{\begin{equation}} 
\newcommand{\ee}{\end{equation}}
\newcommand{\jpa}{J. Phys. A}
\newcommand{\la}{\langle}
\newcommand{\ra}{\rangle}
%%%%%%%%%%%%%%%%%%%%%%%%%%%%%%%%%%%%%%%%%%%%%%%%%%%%%%%%%%%%%%%%%%%

\begin{abstract}
We study a family of diffusively coupled chaotic maps 
on periodic  $d$-dimensional square lattices. Even and odd sub-lattices 
are updated alternately, introducing an effective {\it delay}.
As the coupling strength is increased, 
the system undergoes a first order phase transition from a multi-stable 
to a synchronized phase. Further  increase in  coupling strength shows 
de-synchronization where the phase space splits into two ergodic 
regions. We argue that the de-synchronization transition 
is discontinuous for piece-wise linear maps, and is continuous for 
non-linear maps which are differentiable.
 \end{abstract}
\maketitle

Synchronization is observed in a wide class of complex systems. 
Typically, it appears when the range of the correlations inside the 
system is of the same order as the system size. 
Few examples of mutual synchronization in complex dynamical systems 
are  flashing fireflies\cite{Buck88}, electronic circuits 
\cite{Heagy} and chemical reactions \cite{Kuramoto}. 
In recent years, synchronization of spatially extended systems 
has drawn considerable interest. In particular coupled map 
lattices (CMLs) \cite{Kaneko}, initially introduced as simple models of 
spatio-temporal chaos, has received a great deal of attention as a model 
of synchronization.  It has been realized that  two different 
replicas of the same CML are coupled directly \cite{Grassberger} 
or through a common external random noise \cite{Livi}, can 
become synchronized for large enough coupling strengths. 
Recently, Ahlers and Pikovsky \cite{Ahlers02}  pointed out that 
synchronization transitions  in one dimensional (1-$d$) CMLs are generically 
in the universality class of the Kardar-Parisi-Zhang (KPZ) model, 
however strongly  nonlinear  maps could be in the universality class 
of Directed percolation (DP). 

From a point of view of general 
statistical theory  synchronization is a non-equilibrium phase transition 
where distinct patches of the CMLs oscillate in phase. The phenomenon 
is similar to the roughening of growing interfaces and 
can  be modeled by multiplicative noise partial differential 
equation\cite{Pikovsky}. This picture is further modified 
by M$\tilde{\rm u}$noz and Pastor-Satorass\cite{Munoz} to incorporate 
first order phase transitions (FOPTs). However, it was not clear if 
the FOPTs observed in these models were just transients of DP. 
In context of non-equilibrium wetting process FOPTs are found in several 
(1+1)-dimensional stochastic models with local interactions\cite{Mukamel}, 
contrary to equilibrium wetting process where phase transitions are not 
possible \cite{Zia} in $1-d$ systems having short-range interactions 
between interfaces and substrates. 
To the best of our knowledge FOPTs in deterministic,
chaotic, extended systems with short range interactions are still lacking, 
although it is known to exist for globally coupled maps \cite{Muller}.

In this Letter we introduce and study a single parameter 
family of piece-wise linear chaotic maps which are diffusively 
coupled on a $d$-dimensional square lattice. A {\it delay} is 
introduced dynamically between sub-lattices by updating them 
alternately.  
One of our interests would be to find if, starting from a random 
initial condition, these sub-lattices synchronize at later times.   
The answer turns out to be 'no', for both very high and low diffusion 
strengths $\epsilon$. However for intermediate $\epsilon$  synchronization 
occurs with the suppression of spatio-temporal chaos. This 
synchronized phase is an {\it unique absorbing state} of the system and 
for piece-wise linear maps (PLMs) the phase boundary is identical with the boundary 
for a stable fixed point.

For $\epsilon=0$, the system visits the whole 
phase-space in time. The phase-space volume decreases as 
$\epsilon$ is increased and at a critical strength 
$\epsilon_A$, the phase-space  suddenly shrinks to a point 
which corresponds to the fixed point of the primitive map.  
This fixed point is stable until  $\epsilon=\epsilon_B>\epsilon_A$. 
Further increase in $\epsilon$ results in 
de-synchronization where the phase space splits  into 
{\em two} ergodic regions about the {\it collective bifurcation points}.  
It may be argued that for PLMs the width of the bifurcation 
is independent of $\epsilon$ and thus synchronization error is 
discontinuous at the critical point. However for non-linear differentiable  
maps the width of the bifurcation vanishes as $\sqrt{\epsilon-\epsilon_{c}}$ 
which results in a continuous transition. We argue that this continuous 
transition is in a different universality class than that of  KPZ and DP 
with critical exponents $\beta=1/2$, $\beta/\nu=0$ and $\gamma=1$ in 
$1$-dimension.

{\it The Model :}  Consider a $d$-dimensional hyper-cubic lattice 
${\cal L}$ of coupled identical maps $f(m,z_{\vec i })$,  where
$z_{\vec i }$ is a real variable at site $\vec{i}\equiv (i_1,i_2\dots i_d)$
with $i_k$ varying from $1$ to $L$. We define  the {\it even} and {\it odd}
sub-lattices  (${\cal L}^e$  and ${\cal L}^o$ respectively) as 
 ${\cal L}^{e,o}= \{\vec i : \sum_k i_k =$ {\it even,odd}$\}$,
and denote $x_{\vec i}$ ($y_{\vec i}$) as the variable of ${\cal L}^e$ 
(${\cal L}^o$). Starting from a random initial configuration, 
$\{x_{\vec i}\}$ and $\{y_{\vec i}\}$  are updated alternately as   
 \begin{eqnarray}
 x_ {\vec i}^{t+1} &=& (1-\epsilon) f(  x_ {\vec i}^{t}) 
   +\frac{\epsilon}{2d} \sum_{{\vec j}\in {\cal N}_{\vec i} } 
     f(  y_ {\vec j}^{t}),\cr
  y_ {\vec i}^{t+1} &=& (1-\epsilon) f(  y_ {\vec i}^{t}) 
  +\frac{\epsilon}{2d} \sum_{{\vec j}\in {\cal N}_{\vec i} } 
    f(  x_ {\vec j}^{t+1}),
\label{eq:XY}
\end{eqnarray}  
where  ${\cal N}_{\vec i}$ is a set of $2d$ nearest neighbors of 
$\vec i$ and $\epsilon$ is the coupling strength, can be seen as a 
diffusion constant. Equivalently, in the first half unit of time 
$\{x_{\vec i}\}$ are updated while $\{y_{\vec i}\}$ are at rest 
and the reverse happens in second half.  We will see later that the  
{\it delay } which is introduced dynamically between sub-lattices 
is responsible for a complete synchronization of the system. 
Note that periodic boundary configuration in all $d$ dimensions 
are used throughout.

  Synchronization occurs when the difference between $z_{\vec i}$ and its 
neighbors vanish at all sites as  $t \to \infty$. Thus, the order 
parameter can be defined as  $\phi=\langle\phi^t \rangle$ where
\begin{equation} 
\phi^t = \frac{1}{2dL^d} \sum_{\vec i \in {\cal L}} 
\sum_{\vec j\in {\cal N}_{\vec i}} 
|z_{\vec i}^t-z_{\vec j}^t|,
\end{equation} 
and steady state average is taken over time and
realizations. Obviously $\phi$ vanishes in the synchronized phase and 
in the unsynchronized phase $\phi>0$. A trivial synchronized phase would 
correspond to the stable fixed point of the CLM, $i~e$, 
$\{z_{\vec i}= z^*\}$. It is 
easy to see from Eq. (\ref{eq:XY}) that $z^*= f(m, z^*).$
For chaotic CLMs without {\it delay} the largest Lyapunov exponent is 
independent of $\epsilon$ and is the same as the Lyapunov exponent of 
primitive map which is positive. Hence, a fixed point solution 
$\{z_{\vec i}= z^*\}$  is unstable for any $\epsilon$. 
With a {\it delay}, however, it can become negative in a region 
$\epsilon_B \le \epsilon \le \epsilon_A$. For 
PLMs $\epsilon_{A,B}$ corresponds to the boundary of linearly 
stable region which can be calculated as follows.  

\begin{figure}
 \includegraphics[width=7.0 cm]{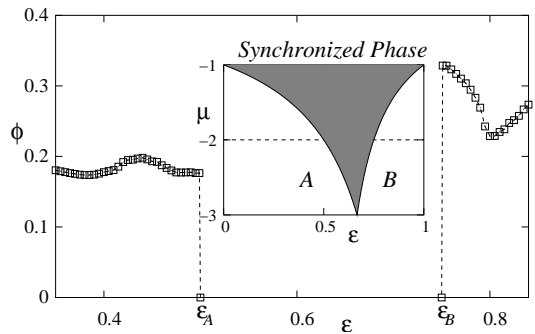}
 \caption{\label{fig:phase} 
Numerically obtained $\phi$ is plotted against $\epsilon$ for 
tent map ($\mu=-2$).
The inset shows the phase diagram in 
$\epsilon$-$\mu$ plane.} 
%The fixed point $z_i=z^*$ is stable in the 
%shaded region, bounded by $\mu=(1-\epsilon)^{-1}$ and 
%$\mu=(2\epsilon-1)^{-1}.$ For piece-wise liner maps these phase boundaries are 
%identical with that of synchronized phase.} 
\end{figure}

  Let us take the initial state to be close to the fixed-point, $i.e.,$
$x_{\vec i}= z^*+\delta x_{\vec i}$ and $y_{\vec i}= z^*+\delta y_{\vec i}.$
In Fourier-space the evolution of $\delta x_{\vec i}$ and 
$\delta y_{\vec i}$  reads as,
\begin{equation}
\left( \begin{array}{c} \delta x_{\vec k} \\ 
\delta y_{\vec k} \end{array}
\right)^{t+1} =
\left(\begin{array}{cc}\tilde\mu& R_{\vec k}\\ 
\tilde\mu R_{\vec k} &
 \tilde\mu+ R_{\vec k}^2\end{array}\right)
\left(\begin{array}{c} \delta x_{\vec k} \\  \delta y_{\vec k} 
\end{array} \right)^t,\label{eq:mat}\end{equation}
$$
{\rm where~} R_{\vec k} = \frac{\epsilon\mu}{2d}
\sum_{\vec r \in {\cal N}_{\vec 0}} 
e^{i{\vec k}.{\vec r}}, ~\mu=f^\prime(m, z^*) 
$$
and $\tilde\mu=\mu(1-\epsilon).$  Let  $E_\pm$ denote the eigenvalues 
of the matrix defined in Eq. (\ref{eq:mat}). From the stability 
requirements $|Max({\cal R}(E_+), {\cal R}(E_-))| <1$,  

we find $\epsilon_A= 1-1/\mu$ and  $\epsilon_B= (\mu+1)/(2\mu)$ which 
are drawn in the inset of Fig. \ref{fig:phase}  as a phase boundary for 
the synchronized phase.  For simplicity, it is assumed here that  primitive 
maps have only single nonzero fixed point $z^*$. One can further 
generalize it to maps with more fixed points.

%----------------------------------------------
%----------------------------------------------
   To find out the behavior of $\phi$ close to these transitions we 
first restrict ourselves to one dimension and  study a specific 
single parameter family of maps : 
\begin{equation}
f(m,z) = \left \{ \begin{array}{ll}
m z/(m-1) & z<a\\
m(1-z)  &  z \ge a 
\end{array}\right. , 
\label{eq:map}
\end{equation}
where $a = 1-1/m$. This piece-wise linear mapping of $[0,1]$ onto itself 
is everywhere expanding for $m>1$, and thus chaotic, 
with an invariant density uniform on $[0,1]$.  
A particular example of this family with $m=2$ is known as {\it tent map}. 
Note that the fixed point is $z^*=m/(m+1)$.
\begin{figure}
 \includegraphics[width=7.0 cm]{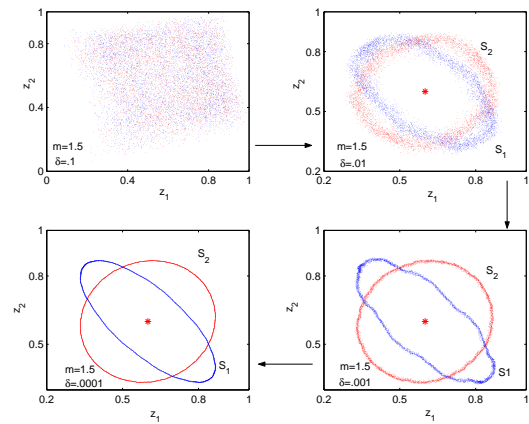}
 \caption{\label{fig:multi} This figure shows how the phase space 
changes in $z_1$-$z_2$ plane as $\epsilon\to \epsilon_A$;  
equivalently $\delta\to 0$. For large $\delta$, the phase space is 
identical for  two different initial configurations $S_1$
and $S_2$. However, dynamically different shapes
are generated as  $\delta\to 0$. The 
symbol '$*$' represents the fixed point $z^*=.6$. }
\end{figure}

{\it Synchronization :}
Let us first discuss the transition from  the unsynchronized 
phase $A$ (inset of Fig. \ref{fig:phase}) to the synchronized phase. 
Close to the transition we take $\epsilon= \epsilon_A-\delta$ 
and find that the system  become multi-stable as $\delta \to 0$, $i.e.,$ 
there are large number steady states and one out of them is chosen by the 
system depending on the initial configuration. Multi-stability  has been 
discussed earlier in the context of CMLs \cite{Chate99}, and 
systems of delayed differential equations\cite{Ikeda}. 
It may be argued that the multi-stability is extensive, $i.e.,$ the 
number of attractors grow exponentially with the system size. 
Thus any statistical average has to be taken over large number of 
independent realizations, which resticts us to simmulate large 
systems. We carried out numerical simmulations for $L=1024$ and $m=2$ 
({\it tent map}) and find that  $\phi$ vanishes  discontinuously at 
$\epsilon_A=0.5$ (see Fig. \ref{fig:phase}).  To confirm that it is 
a true first order transition, not just a transient 
effect, we monitor the phase space of every neighboring pair of 
co-ordinates as $\delta \to 0$. For example in Fig. \ref{fig:multi} 
we demonstrate how the phase space changes in $z_1$-$z_2$ plane.  
Every other pair of neighboring co-ordinates show similar changes. 
In practice, no  noticable change is observed in the phase space when 
$\delta<10^{-4}$ and then suddenly the fixed point $z_1=z^*=z_2$ appears 
at $\delta=0$. In other words, when $\delta\approx 0$, 
we have $|z_i-z_{i+1}|>0$ for every realization and thus $\phi$ has 
a jump at the critical point $\epsilon_A$.

\begin{figure}
 \includegraphics[width=7.0 cm]{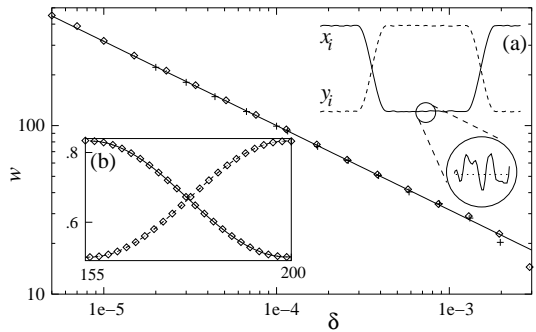}
 \caption{\label{fig:kink} A typical steady state profile of odd (solid-line) 
and even (dashed line) sub-lattices are shown in inset (a). 
Inset (b) compares the kink profile obtained from simulations (points) with 
Eq. (\ref{eq:kink}). In the main figure, width of the 
kink $w$ is plotted against $\delta$ in log scale for  $m=2$ 
and $2.5$. The slope of the solid line is set to $-1/2$.} 
 \end{figure}

{\it De-synchronization :}  The synchronized state persists 
up to $\epsilon=\epsilon_B$ where phase space splits into 
{\it two} disconnected ergodic regions. In the new phase 
$B$ shown in the inset of Fig. \ref{fig:phase}, the system fluctuates 
about  its {\it collective bi-bifurcation points} 
$x_{i}= x^*$ and $y_i= y^*$. Using Eq. (\ref{eq:XY}) we get   
\begin{eqnarray}
x^*&=& (1-\epsilon) f(x^*)+\epsilon f(y^*)\cr
y^*&=& (1-\epsilon) f(y^*)+\epsilon f(x^*),
\label{eq:bi}
\end{eqnarray}
 which can be solved for the family of maps defined in Eq. (\ref{eq:map}) 
as $x^*=\alpha_\pm$ and corresponding $y^*= \alpha_\mp$, with 
$\alpha_\pm=\frac{m^2(2\delta+1)\pm (2\delta m+1)}{m^2(2\delta+1) 
+m(4\delta +1)}$ and $\delta=\epsilon-\epsilon_B$. 
Depending on the initial configuration, different parts of the sub-lattices 
are then attracted to $\alpha_+$ or $\alpha_-$ with kink-like interfaces 
(inset (a) of Fig. \ref{fig:kink})
separating them. It will be shown later that the width of such a kink $w$,  
diverges as $1/\sqrt\delta$ as $\delta\to 0$. Thus,  stable kinks can 
not be generated when $\delta$ is ${\cal O}(L^{-2})$
and we have $\phi= \alpha_+-\alpha_-$. Clearly the jump in $\phi$ 
at the critical point $\epsilon_B$ is $\Delta=2/(m^2 +m)$. Note that in the 
other limit, $i.e.$, when $L \to \infty$ before $\delta \to 0$, we have a 
slightly different $\phi$. In this case the large systems  
will generate an  average density of kinks, say $\rho$. Immediately,  
we have $\phi= \Delta( 1- \frac{\rho{\cal A}}{w\Delta})$, where 
${\cal A}$  is the area bounded by an even and an odd kink. 
Nevertheless the transition is discontinuous.  

    The profile of a kink can be calculated as follows. 
Let us assume that it  starts  at site $k$ with $z_{k}=\alpha_-$ and  
$z_{k+1}=\alpha_+$. In steady  state $z_{k+2}$  has two solutions : 
$\alpha_-$ and  $\alpha = \frac{m^2(2\delta+1) +(2\delta m-1)}
{m^2(2\delta+1) + m(4\delta +1)}$. 
It is obvious that the first solution, $z_{k+2}=\alpha_-$, does not 
generate a kink. Thus the steady state profile of a kink can be obtained as 
\begin{equation}
\left( \begin{array}{c} z_{k+i+1}\\ z_{k+i+2}\end{array}\right) =
 \left( \begin{array}{cc}0&1\\-1 &-2\cos(\theta) \end{array}\right)^i
\left( \begin{array}{c} \alpha_+\\ \alpha\end{array}\right),
\label{eq:kink}
\end{equation}
where $\theta=2\tan^{-1}(\sqrt{\delta/\epsilon_B})$.  
Explicit form of $z_{k +i}$  is lengthy to present here, 
rather we compare numerically obtained $z_{k+i}$ with Eq. (\ref{eq:kink}) 
in the inset (b) of Fig. \ref{fig:kink}. When site $i$ is far from 
$k$ and $\delta\approx 0$  one can write  an approximate
solution as $z_{k+2i}=\alpha_- + \Delta_i$ 
and $z_{k+2i+1}=\alpha_- -\Delta_i$, where $\Delta_i=  
\frac{i^2 \delta}{2m(m+1)}$. The width of the kink is thus 
$w=2n$, such  that $\Delta_n = \alpha_+ - \alpha_-$.
Clearly, $w$ diverges as $1/\sqrt\delta$. In Fig.\ref{fig:kink} we
have plotted $w$ obtained  from numerical simulations 
versus $\delta$ in $\log$-scale, which confirms this power-law. 
%-------------------------------------------------
%-------------------------------------------------

{\it Higher Dimensions :} The phase diagram for coupled PLMs 
in higher dimension is the same as shown in the inset of 
Fig.\ref{fig:phase}. Close to the transition point $\epsilon_A$, phase 
$A$ is mutistable and shows sudden shrink in phase space volume. The 
desynchronization transition occurs at $\epsilon_B$. But unlike one 
dimension, the kink-type solutions in phase $B$ are no more stable 
and hence jump in the orderparameter is $\Delta=2/(m^2 +m)$. 
Numerical simulations  in $2$ and $3$ dimensions confirms this finding.

{\it Other Maps :} How do non-linear maps behave when a
{\it delay} is introduced between sub-lattices ? Clearly the
phase boundaries for synchronized phase  strongly depends on 
corresponding primitive maps.  First let us consider 
the {\it logistic map} :  $f(x)=4x(1-x).$ The fixed point 
of this map $z_i = z^*=3/4$ is linearly stable for $1/2<\epsilon<3/4$.    
However numerical simulation shows that the synchronization 
occurs at $\epsilon_A> 1/2$. Close to the transition point phase $A$ is 
multistable and hence we have a first order transition similar 
to that of the {\it tent map}. The de-synchronization is found 
to occur exactly at $\epsilon_B$, where the linear-stability of the 
fixed point breaks down. Unlike the tent maps here we do not have 
fluctuations about  the bifurcation points. But the kinks are 
unavoidable. Using Eqs. (\ref{eq:bi}) and  (\ref{eq:kink}) one can 
obtain the bifurcation points as $x_i=\alpha_\pm$ and  $y_i= \alpha_\mp$, 
where $\alpha_\pm= \frac{8\delta+3\pm2\sqrt\delta(8\delta+3)}{4(1+4\delta)}$.
Clearly   $\Delta = \alpha_+ -\alpha_- $  vanishes as $\sqrt{\delta}$ 
indicating that the transition is continuous. Critical exponents of 
this continuous transition can be deduced using following arguments.   

First, note that fluctuation of 
order-parameter comes about from the  variation in  density of kinks 
which fluctuates about a  mean density, say $\rho $. Density distribution 
about $\rho$ can be assumed to be {\it normal}, which   gives 
$\phi \sim \sqrt\delta (1-\frac{{\cal A}\rho}{w\Delta})$,  where 
${\cal A}$  is the area bounded by an odd and an even kink, and 
$w$ is the width of the kink. Since $\phi$ is independent of $L$ and 
proportional to  $\sqrt\delta$, we have critical exponents
$\beta=1/2$ and $\beta/\nu =0$. One can further define ``susceptibility'' 
as  $\chi= (\langle\phi^2\rangle - \langle\phi\rangle^2)/L$. 
Now $\chi \sim \delta r^2/L$, and hence $\gamma=1$. It is evident that 
the critical dimension $d_c=1$. 

We have also done numerical simulations on  {\it sine} ($f(x)=\sin(\pi x)$) 
and {\it cubic} ($f(x)=\sqrt{27}x(1-x^2)/2)$  maps in one dimension, 
which show that the synchronization transition  is discontinuous, whereas 
de-synchronization occurs continuously with the same critical exponents 
as that of the {\it logistic } map. 

To get an insight 
why de-synchronization is continuous  for certain class of 
maps  we investigate  coupled  {\it power-law} maps: $f(z) = 
1- |2z-1|^b$,  where $b=1$ and $2$ corresponds  to the {\it tent} and 
the {\it logistic} maps respectively. Thus, by tuning $b$ it is possible 
to study  how the nature of transition changes from being first to 
second order. Our numerical simulations suggests that  for $b>1$ the 
de-synchronization transition is continuous and belongs to the same 
universality class as that of the {\it logistic } map, whereas for  
$1/2<b\le 1$  it  is discontinuous. Note that there are {\it two} non 
zero fixed points  for $b<1/2$ and present analysis need to be modified 
for their discussion . 
Basically, synchronization error is the average difference between 
bifurcation points $x^*$ and $y^*$, which vanishes 
at $\epsilon_B$ where $x^*=1/2=y^*$. 
Just above $\epsilon_B$ this difference is proportional 
to the difference between the left and right slope of $f(z)$  
at $z=1/2$. For $b<1$, since these maps are not differentiable at $z=1/2$ 
the de-syncronization occurs discontinuously, whereas for $b>1$ the
maps are differentiable and thus we have a continuous transition.
%{\it The nature of transition in delayed coupled maps are thus 
%dictated  by the differentiability properties of the primitive maps}.  

%-------------------------------------------------
\begin{figure}
 \includegraphics[width=7.0 cm]{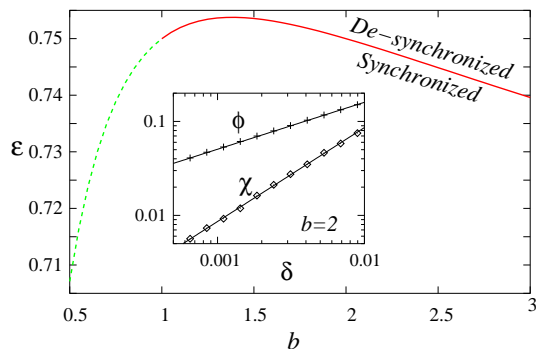}
 \caption{\label{fig:log} Phase diagram of de-synchronization transition is 
shown in  $\epsilon$-$b$ plane. The nature of transition changes 
from being discontinuous (dashed line) to continuous (solid line) at $b=1$.
In the inset, $\phi$ and $\chi$ versus $\delta$ is plotted in 
log-scale and the solid lines are drawn with slope 
$\beta=1/2$ and $\gamma=1$ respectively for comparision.
}
\end{figure} 
%-------------------------------------------------

      A few comments are in order.  First, the synchronized phase 
discussed here is different from earlier studies, where CMLs 
remain chaotic  in the synchronized phase for both, when they are 
coupled directly \cite{Grassberger} or through noise \cite{Livi}. 
In our model the spatio-temporal chaos are suppressed 
and thus the synchronized phase is a completely ordered phase. 
The spatial correlations even persist in de-synchronized phase $B$. 
Second,  one can 
define a different order parameter called ``magnetization'' 
$M=\langle m^t\rangle$,  where $m^t = \sum_i sign(z_i^t-z^*)$. 
Then the desynchronyzation transition can be thought as a 
transition from fully ferromagnetic to fully antiferrmagnetic 
phase, which can be either first  or second order. Note that the 
continuous transition obtained here is in a different universality 
class than that of zero temperature Ising transition in 
one dimension.

 {\it Conclusion :}  In conclusion, we study a single parameter 
family of coupled chaotic maps by introducing a {\it delay} 
between sub-lattices. We show that these systems undergo a first order 
transition  to a synchronized phase when the coupling parameter 
is varied. The transition occurs as the system enters from a 
multiply stable region to a  single `collective fixed point' in phase 
space. A second transition occurs as this fixed point become unstable 
and the phase space breaks up into two ergodic regions about the 
collective bifurcation points.
 Our analytical results show that 
the de-synchronization transition is discontinuous in one and higher 
dimensions when the primitive maps are not differentiable. 
Differentiable maps also show a first-order synchronization transition,  
whereas the de-synchronization occurs continuously with critical exponents 
$\beta=1/2$, $\gamma=1$, and $\nu^{-1}=0$. We argue that the 
de-synchronization 
transition discussed here can also be thought as a
transition from a fully ferromagnetic to a fully anti-ferromagnetic 
phase. These results establish a true non-equilibrium first order transition 
in one dimensional dynamical systems with short range interactions.
 
{\it Acknowledgments :} We thank D. Mukamel and E. Levin for 
fruitful comments and discussions.

\end{document}